# A NOVEL X-AXIS TUNING FORK GYROSCOPE WITH "8 VERTICAL SPRINGS–PROOFMASS" STRUCTURE ON (111) SILICON

*Fei Duan[1], Jiwei Jiao[1], Yucai Wang[1], Ying Zhang[1], Binwei Mi[1], Jinpeng Li[1], Jian Zhu[2], Yuelin Wang[1]*

1. State Key Laboratory of Transducer Technology, Shanghai Institute of Microsystem and Information Technology, Chinese Academy of Sciences, China
2. Micro/Nano R&D Center, Nanjing Electronic Devices Institute, China

**ABSTRACT**

A novel x-axis tuning fork MEMS gyroscope with "8 vertical springs-proofmass" structure for Coriolis effect detection is presented. Compared with the common single-plane springs, the 8 vertical springs, symmetrically located at the top and bottom sides, more stably suspend the large thick proofmass featuring large capacitance variation and low mechanical noise. A bulk-micromachining technology is applied to obtain the large proofmass and twins-like dual beams. During the fabrication process, the dimensions of the 8 vertical springs are precisely confined by thermal oxide protected limit trenches (LTs) sidewalls and the extreme slowly etched (111)-planes; therefore a small mismatch of less than 30 Hz is achieved before tuning. Initial test shows a sensitivity of 0.15mV/(deg/s) and rate resolution around 0.1deg/s under atmosphere pressure.

## 1. INTRODUCTION

Planar x-axis gyroscopes have attracted much attention for multiple-axis sensing applications not only in automotive and navigation fields. Low-cost and integrated multiple-axis micro gyroscopes have even attracted the handset manufacturers' attention. The monolithic integration approach of x-, y-, z-axis gyros is preferable, if possible, to the common 3D-assembly, because of the smaller size, better orthogonality and easier packaging.

Since a planar y-axis gyroscope is easily obtained by rotating the x-axis one by 90 degree in the mask layout, and as well a z-axis gyroscope can be easily done with only in-plane motions, the multiple-axis gyroscope integrated on a single wafer can be realized if an x-axis gyroscope is developed. However, it seems a challenge to detect the proofmass' Coriolis-force-led out-of-plane motion in an x-axis gyroscope. Till now much fewer x-axis gyroscope prototypes have been reported [1,2] than their vertical counterparts. Commonly used SOI material [1,2] limits the inertial masses that may raise the noise floor accordingly. Single proofmass and single-side-beam structure[2] possibly result in extra cross-talks. Vertical springs-proofmass structure was implemented in precision accelerometers[3,4], but aligned bonding and doping introduced stress control are strictly required.

In this paper, we present a novel x-axis gyroscope with "8 vertical springs-proofmass" structure on (111) silicon, which can effectively restrain the cross-axis coupling to harsh circumstance and process imperfection. Bulk micromachining processes, compatible with the processes of z-axis gyro developed by us[5], are applied to fabricate this gyro. This technology facilitates the fabrication of x-axis microgyroscopes with large capacitance variation, controllable damping, small cross-talks and low mechanical noise.

## 2. DESIGN AND SIMULATIONS

The proposed tuning fork gyroscope consists of two connected identical oscillating frames. When the gyro is working, contrary movements of these two frames eliminate the disturbance from the axial acceleration as well as double output amplitude. A detailed cross-section view is shown in Fig.1. As the device is driven along y-axis by electromagnetic actuation and an external angular rate is applied about x-axis, the proofmasses move vertically (along z-axis) owing to the Coriolis force. This vertical motion is detected by the gap-changing parallel-plate capacitors between the silicon plate and glass substrate. The decoupled driving and sensing modes is designed to prevent unstable operation due to mechanical coupling.

Fig.2 schematically shows the Corilis-effect-detecting resonator inside the gyro. The inner resonator, which is suspended inside a movable rigid decoupling frame in the gyro, consists of 8 suspension springs and a central proofmass. The 8 vertical springs, actually 4 twins-like dual-beam pairs, are designed in quasi crab-leg shape for smaller size and placed symmetrically about the proofmass.

*Fei Duan, Jiwei Jiao, Yucai Wang, Ying Zhang, Binwei Mi, Jinpeng Li, Jian Zhu, Yuelin Wang*
*A NOVEL X-AXIS TUNING FORK GYROSCOPE WITH "8 VERTICAL SPRINGS-PROOFMASS" STRUCTURE ON (111) SILICON*

The vibration modes of such "8-1 type" resonator ("8 vertical springs-proofmass" structure) are investigated, in comparison to the common "4-1 type" (4 single-plane-beams with one proofmass) one of the same parametric geometric dimensions, which are listed in TABLE I, with a finite-element approach. The three-dimensional (3D) solid models do not contain the etching holes in the proofmass for simplified meshing; however, the simulations take into account these effects by using effective equivalent mass holes.

TABLE I
LIST OF SIMULATION GEOMETRIC DIMENSIONS

| GEOMETRIC PARAMETER | DIMENSION DATA |
|---|---|
| Vertical springs (dual beam) | 1800×50×30 (μm) |
| Seismic mass | 5300×2000×300 (μm) |
| Effective equivalent hole | 1600×800×300 (μm) |
| Single-plane beam | 1800×50×60 (μm) |

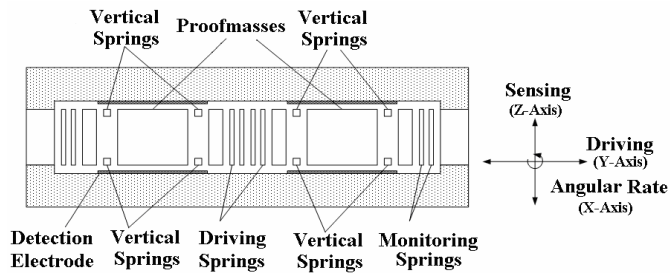

Fig.1 Schematic cross-section view of the gyroscope

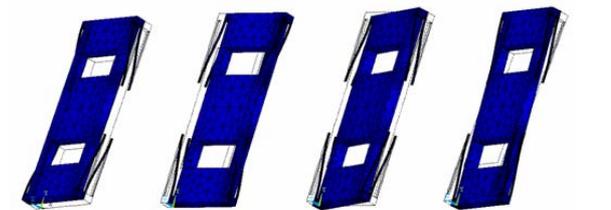

$1^{st}$(2386Hz)   $2^{nd}$(3419Hz)   $3^{rd}$(5098Hz)   $4^{th}$(6010Hz)
(a) $1^{st}$-$4^{th}$ vibration modes of "8-1 type"

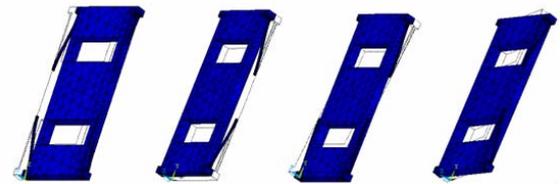

$1^{st}$(3441Hz)   $2^{nd}$(4263Hz)   $3^{rd}$(4522Hz)   $4^{th}$(8547Hz)
(b) $1^{st}$-$4^{th}$ vibration modes of "4-1 type"

Fig.3 First four vibration modes of "4-1 type" and "8-1 type"

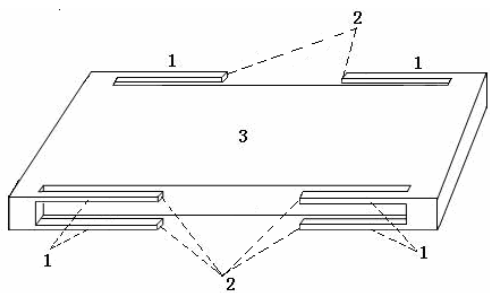

**1**. Vertical springs (dual beam pairs)
**2**. Anchors
**3**. Proofmass

Fig.2 Schematic inner resonator of "8 vertical springs-proofmass" structure

The early four vibration modes and frequencies of the two types are demonstrated in Fig.3. The translational vertical motion, acting as the usable mode for out-of-plane detection, is the fundamental mode for the "8-1 type". However, for the "4-1 type", it is the $2^{nd}$ mode, close to the $3^{rd}$ disturbing torsion mode. The much less frequency difference between the usable mode and neighboring mode for the "4-1 type"($2^{nd}$-$3^{rd}$, 259 Hz), than that for "8-1 type"($1^{st}$-$2^{nd}$, 1033 Hz), implies greater elastic coupling among translational (x-, y-, z-) and rotational ($\Phi x$, $\Phi y$, $\Phi z$) aspects[6]. It is evident that the twins-like dual beams accompanying with large thick proofmass, as the "8-1" type, is more resistent of extra cross coupling.

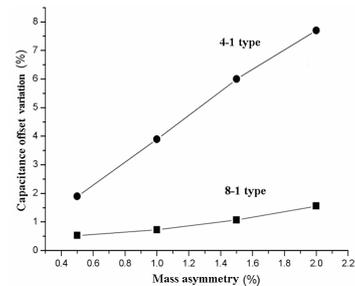

Fig.4 FEM simulations showing capacitance offset variation versus mass asymmetry (0.5-2%) of "4-1 type" and "8-1 type", respectively

The case concerning the manufacturing variation is also considered. Typical situation of asymmetrical proofmass, which possibly becomes observable for large thick mass plate, results in the capacitance offset variation. Fig.4 presents the FEM simulation results showing the capacitance offset variation versus mass asymmetry ranging from 0.5% to 2% and indicates more resistency of mass asymmetry due to manufacturing variation for the "8-1 type".

The x-axis gyroscope is designed to work under the atmosphere pressure. Considering DRIE conditions, the $50\times50$ μm² air damping holes impenetrate the proofmass to reduce the squeeze film air damping. Air damping of the perforated thick plates is estimated through the Modified Reynolds' Equation[7]. Accordingly the quality factor in the sensing direction remains 67.

## 3. FABRICATION

Fig.5 shows the main fabrication processes, starting with a 350μm-thick low-resistivity (111)-silicon wafer.

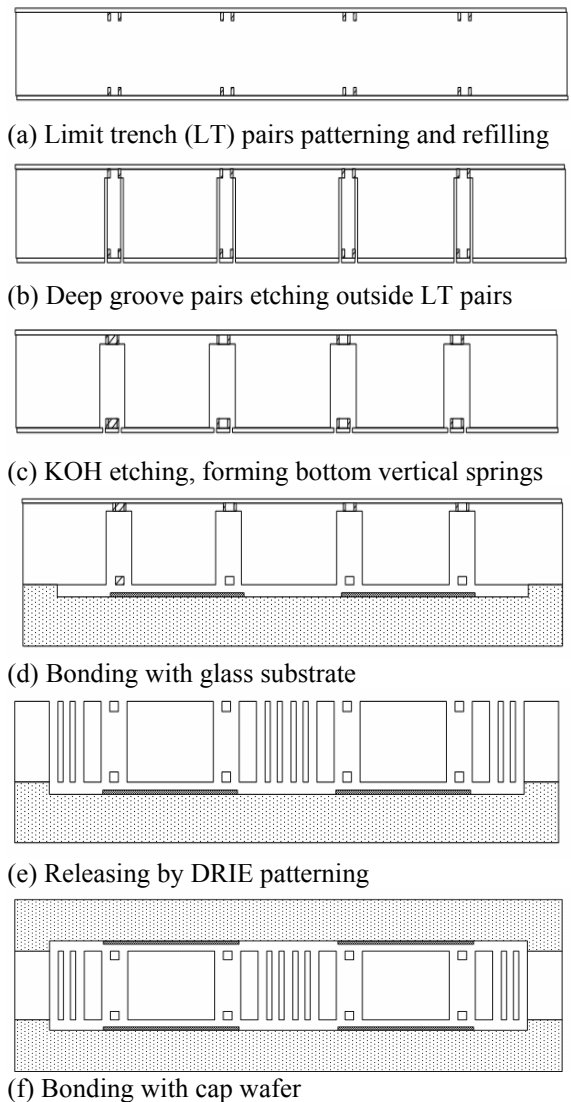

(a) Limit trench (LT) pairs patterning and refilling

(b) Deep groove pairs etching outside LT pairs

(c) KOH etching, forming bottom vertical springs

(d) Bonding with glass substrate

(e) Releasing by DRIE patterning

(f) Bonding with cap wafer

Fig.5 Fabrication flow

(a) 30μm deep limit trench (LT) pairs are formed symmetrically at both sides of the wafer to define the vertical spring areas, and refilled with thermal SiO$_2$. There are 4 LT pairs around each proofmass area.

(b) From the bottom side, deep groove pairs, slightly offset outside of the LT pairs, are etched with DRIE till carefully over the low end of the top LTs. Both LTs and groove pairs are longitudinally along <110> direction.

(c) Next step is KOH etching. Unprotected (110)-plane sidewalls of the groove pairs are fast etched away till the lateral etching fronts meet. SiO$_2$ protected LT sidewalls and the extreme slowly etched (111)-plane precisely confine the dimension of the 8 bottom vertical springs, as well as the thickness of the top vertical springs.

(d) The (111) wafer is then bonded with a Pyrex glass, which is covered with detection electrodes in beforehand made 5um recesses for gap-variational sense.

(e) From the top side, DRIE is applied to release top vertical springs, proofmasses, driving/monitoring springs, and damping holes, followed by the second bonding with a cap glass wafer (f).

A double-layer composite mask composed of thermal oxide and PR (photo resist) is applied during the releasing step to protect the back-side vertical springs.

## 4. EXPERIMENTAL RESULTS

Fig.6 is the top SEM view of a gyroscope with "8 vertical springs-proofmass" structure before the second bonding. Fig.7 is the close-up view of the released vertical top and bottom springs. The fabricated device was packaged in air and experimentally evaluated using our self-developed testing circuit.

The resonance frequencies for drive and sense modes are 3998 and 4020 Hz, respectively, as shown in Fig.8, with a mismatch of 22Hz before tuning.

Fig.9 (a) shows the measured input excitation signal for the drive mode and (b) shows the corresponding output response picked by the interface circuit. A bandwidth of at least 5Hz is attained.

Fig.10 shows the measured output as a function of angular rate within the range of ±200deg/s corresponding to a sensitivity of 0.15mV/ (deg/s), which is tested by fastening the device (including the interface system) on arate table (Ideal Aerosmith Model 1280). Our efforts on damping control and mode tuning seem promising to improve the sensing Q-factor and device's sensitivity further.

Fig.11 shows the frequency response of the gyro to a 10deg/s, 2Hz angular rotation. The amplitude of the angular-rate-responded peak is nearly 40dB larger than the noise floor, which indicates the noise equivalent rate resolution around 0.1deg/s.

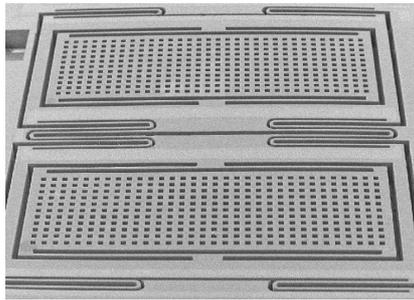

Fig.6 Top view of the x-axis gyroscope

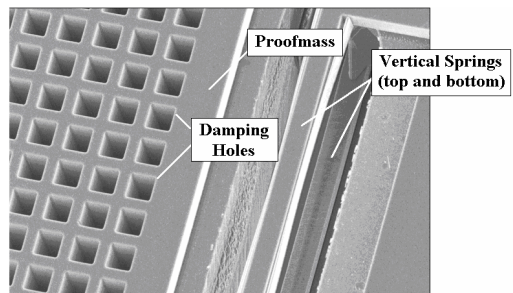

Fig.7 Close-up view of the vertical springs

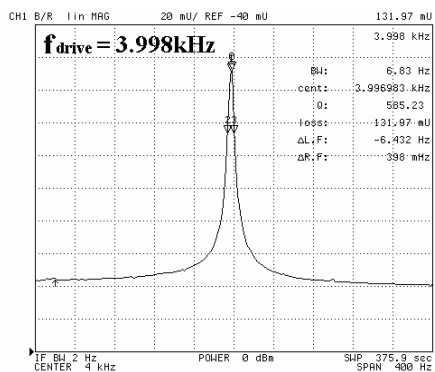

(a) Drive mode

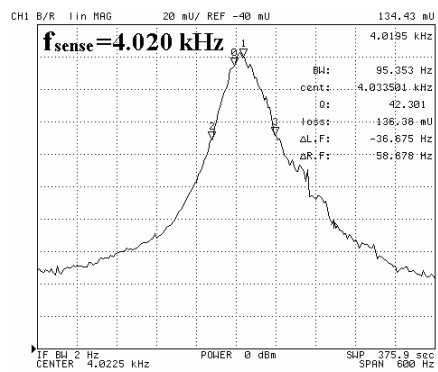

(b) Sense mode

Fig.8 Resonance characteristics of drive mode and sense mode

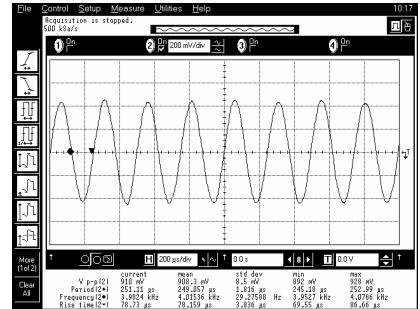

(a) Excitation signal

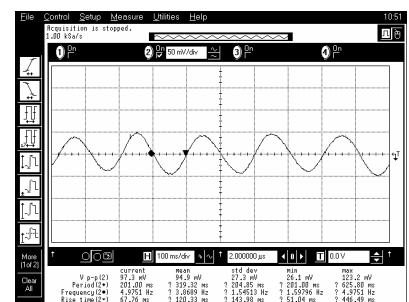

(b) Output response signal
(responding to 300deg/s, 5Hz angular rate)

Fig.9 Excitation signal and output response signal in time domain

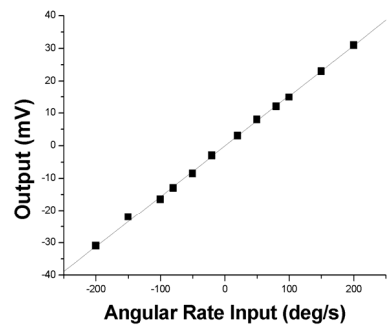

Fig.10 Output versus angular rate input within ± 200 deg/s

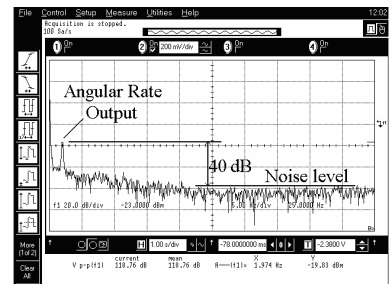

Fig.11 Frequency response to a 10deg/s, 2Hz angular rate input

## 5. CONCLUSION

A novel x-axis tuning fork microgyroscope on (111) silicon is designed, fabricated and tested. The inside "8 vertical springs-proofmass" structure restrains the cross talks by attaining stable Coriolis-effect detection and the large proofmass achieves large capacitance variation and low mechanical noise. A bulk micromachining technology is successfully applied to obtain the x-axis gyro. The fabrication with precise determination of beam dimensions leads to easy mode matching. Initial test shows a sensitivity of 0.15mV/(deg/s) and rate resolution around 0.1deg/s under atmosphere pressure. The fabrication technology of this x-axis gyro is fully compatible with that of our developed z-axis gyro, which reveals the potential of realizing 3D integrated microgyroscopes in a single wafer.

## ACKNOWLEDGEMENT

This research is supported by the National High Technology Research and Development Program of China (863 Program) under Grant no.2005AA404020. This support is gratefully acknowledged.